\documentclass[prb,
showpacs, prbprintnumbers, amsmath, amssymb, nodata]{revtex4}

\def\Ref#1{(\ref{#1})}
\newcommand{\be}{\begin{equation}}
\newcommand{\ee}{\end{equation}}
\newcommand{\bn}{\begin{eqnarray}}
\newcommand{\en}{\end{eqnarray}}
\usepackage{epsfig}
\usepackage{graphicx}
\usepackage{dcolumn}
\usepackage{bm}
\usepackage[english]{babel}

\begin{document}


\title{ Relaxation time scales in collective dynamics of liquid
alkali metals
        }

\author{\firstname{Anatolii~V.}~\surname{Mokshin}}
\email{mav@theory.kazan-spu.ru} \affiliation{Department of
Physics, Kazan State Pedagogical University, 420021 Kazan, Russia}

\author{\firstname{Renat~M.}~\surname{Yulmetyev}}
\email{rmy@theory.kazan-spu.ru} \affiliation{Department of
Physics, Kazan State Pedagogical University, 420021 Kazan, Russia}

\author{\firstname{Peter}~\surname{H\"anggi}}
\affiliation{Department of Physics, University of Augsburg,
D-86135 Augsburg, Germany}


\date{\today}

\begin{abstract}
In this paper the investigation of the dynamical processes of
liquid alkali metals is executed by analyzing the time scales of
relaxation processes in liquids. The obtained theoretical dynamic
structure factor $S(k,\omega)$ for the case of liquid lithium is
found to be in excellent agreement with the recently received
inelastic X-ray scattering data. The comparison and interrelation
with other theories are given here. Finally, an important part of
this paper is the confirmation of the scale uniformity of the
dynamic processes in liquid alkali metals predicted by some
previous molecular dynamic simulation studies.
\end{abstract}

\maketitle


\section{\label{intr} Introduction}

The dynamic structure factor $S(k,\omega)$ is an experimentally
measured term, containing information about the processes in a
liquid with long- and short-time scales. It can be used to judge
on the microscopic behavior in a system on the basis of its
spectra, obtained by means of Inelastic Neutron Scattering (INS)
\cite{Balucani_Zoppi,Copley} or Inelastic X-ray Scattering (IXS)
\cite{Burkel}. As for simple liquids, at present a great amount of
experimental data of $S(k,\omega)$ has been accumulated, in
particular, for liquid alkali metals. These data indicate legibly
the presence of the collective propagation excitations beyond the
hydrodynamic region. The characteristic feature of liquid alkali
metals is a triple-peak shape of $S(k,\omega)$ lasted to $k \sim
0.8 k_m$, where $k_m$ corresponds to the first maximum of the
static structure factor $S(k)$. Moreover, the frequency of the
side peak achieves its maximum at $k \sim 0.55 k_m$. The
propagation of these high-frequency waves cannot be obtained
within a hydrodynamic treatment, therefore, they are related in
some works to the so-called kinetic collective excitations. The
impossibility to describe these microscopic phenomena and,
therefore, to reproduce qualitatively the experimental
$S(k,\omega)$ by means of ordinary hydrodynamic equations led to
the development of other theoretical models and approaches.

One of the simplest and perhaps the earliest modeling approaches
is the so-called viscoelastic theory. It allows one to obtain the
central quasi-elastic line as well as two inelastic peaks
symmetrically located around $\omega=0$ for mesoscopic
space-frequency region. However, as shown in Refs.
\cite{Bodensteiner,Tullio0}, this model can not be used for the
exact reproduction of the experimental spectral shapes of
$S(k,\omega)$ (see, for instance, the cases of liquid cesium and
lithium in Refs. \cite{Bodensteiner,Tullio0}). Therefore, in Ref.
\cite{Tullio0} the double-scale model for the viscous relaxation
process with fast and slow time scales was tested, and as a result
a good agreement with the IXS experimental data for the dynamic
structure factor was received. Recently the similar approach was
also applied for the description of relaxation processes in
H-bonded liquids \cite{Angelini,Balucani1}. The existence of two
time scales in this model reflects the presence of physically
different decay mechanisms. A faster process is hypothetically
associated with interactions between an atom and the ``cage'' of
its nearest neighbors, and a slower one is identified with the
well-known structural $(\alpha-)$ process. However, relaxations of
both processes are approximated by exponential dependencies. In
recent works the viscoelastic model has been improved by means of
the Markovian closure on the next relaxation level of Zwanzig-Mori
hierarchy \cite{Cabrillo}, it is equivalent to the exponential
relaxation on this level. It is worth mentioning two others
methods, one of which is related to the extension of the usual
hydrodynamic analytical expressions by modification of
hydrodynamic modes to $k$-dependence (see, for instance, Ref.
\cite{McGreevy}). This method assumes the existence of
non-hydrodynamical additional modes. The second approach is
related to the so-called concept of generalized collective modes,
which was proposed for the investigation of the time correlation
functions (TCF's) beyond the hydrodynamic region \cite{Bruin}. The
key idea of this method consists in the correct choice of the
basic set of dynamical variables.

All these methods are more or less successfully used for the
description of collective dynamics in liquids. They have common
property. Namely, they are actually constructed on heuristic
assumption about the presence of exponential decay (or combination
of exponential decay contributions) in some relaxation processes.
Nevertheless, the transition and imposition of different
relaxation modes in disorder systems can occur even in case of a
concrete relaxation process, that complicates the selection of the
analytical time dependence for the corresponding TCF. This fact is
proved by the successful application  of different mode-coupling
theories. On the other hand, this difficulty can be resolved by
means of analysis and comparison of the resulting time scales of
relaxation processes. Therefore, in the present work we suggest
the approach, which allows us to avoid the immediate approximation
of relaxation processes by analytical functions. It is based on
the development of Bogoliubov's ideas about the hierarchy  of
relaxation times in liquids \cite{Bogoliubov}, adapted to the
formalism of time correlation functions.

One of the open problems in studying of liquid state (in
particular, of the microdynamics of simple liquids) is to describe
and understand on a general ground the common features of
different relaxation processes \cite{Angelini}. It is well known,
that the dispersion of the side (high-frequency) peak of dynamic
structure factor is the same for all alkali metals. Moreover, it
is also valid in case of more complex systems, for example, for
liquid alloys \cite{Bove}. Then the following questions arise: Is
the origin of relaxation processes the same for liquid systems
with the similar features? Can the unified description be applied
to these systems? As for the group of melting alkalis, it has been
indicated in Ref. \cite{Balucani_PRB} that both the equilibrium
and the time-dependent correlations can be cast in a properly
scaled form for all the the alkali metals. Further, it was
justified by \textit{ab initio} molecular dynamic studying in Ref.
\cite{Balucani_PRB} too. Experimental confirmation of this result
was impossible over a long period particularly because of the
difficulties related to the technique of INS due to the deficient
precision of the experimental data. Recently, due to progress in
IXS technique this issue was considered again \cite{Tullio1}. In
this work we present investigations related to the determination
of corresponding scale transitions for liquid systems.

The organization of the paper is as follows. In the next Section,
we describe the theoretical formalism, and the comparison with the
experimental data and other theories is carried out. The
possibility of scale uniformity of dynamical processes in the
group of liquid alkali metals is analyzed and discussed in Section
\ref{scale}. The scale-crossing relations are also presented here.
Finally, we come up with some concluding remarks in Section
\ref{remarks}.

\section{\label{sec:level1} Theoretical formalism}
\subsection{\label{sec:level2} Basic notions}
Let us consider the liquid system of $N$ identical classical
particles of the mass $m$ in the volume $V$ and take the density
fluctuations \be
W_{0}(\textbf{k})=\frac{1}{\sqrt{N}}\sum_{j=1}^{N} e^{i \textbf{k}
\cdot \textbf{r}_j} \label{W0} \ee as an initial dynamical
variable. To construct a some set of dynamical variables necessary
for the description of the evolution of the system we use the
technique of projection operators of Zwanzig-Mori
\cite{Zwanzig,Mori}. It is a formal version of the Gram-Schmidt
orthogonalization process, which allows one to obtain the set of
\textit{orthogonal} variables \bn \textbf{W}(k)=\{W_0(k), W_1(k),
W_2(k), \ldots , W_j(k), \ldots \}. \en They satisfy the condition
$\langle W_j^* W_l \rangle= \delta_{j,l} \langle |W_j|^2\rangle$
and are connected by the following recurrent relation
\cite{Yulm_Khus}: \bn && W_{j+1}(k)=\mathcal{L} W_{j}(k)
-\Omega_{j}^2(k) W_{j-1}(k),\nonumber\\ && j=0,\ 1,\ 2,\ldots;\
W_{-1}(k)=0. \label{RR} \en Here the characteristic of the
corresponding $j$th relaxation process, the so-called frequency
parameter $\Omega_{j}^2(k)$, appears, $\mathcal{L}$ is the
Liouville operator \be \mathcal{L}=-i \left \{
\sum_{j=1}^{N}\frac{\textbf{p}_j \nabla_j}{m} - \sum_{i>j=1}^{N}
\nabla_j\; u(j,i) ({\displaystyle \nabla_p^j - \nabla_p^i}) \right
\} \label{liuvillian} \ee with the momentum of the $j$th particle
$\textbf{p}_j$ and the pair potential  $u(j,i)$. So, if $W_0(k)$
is the density fluctuations, then $W_1(k)$ is the longitudinal
component of the momentum density and so it goes on.

The TCF's for the corresponding dynamical variables are given by
\be M_{jl}(k,t)=\frac{\langle W_j^*(k) e^{i \mathcal{L}_{22}^{(l)}
t} W_l(k)\rangle}{\langle W_j^*(k) W_l(k)\rangle}, \  j,l=1,2,...
\label{TCF} \ee For convenience \textit{normalized} time
correlation functions are used here. The time-evolution operator
of Eq. \Ref{TCF} contains the reduced Liouville operator \be
\mathcal{L}_{22}^{(l)} =\left (1-\sum_{j=1}^l \Pi_j \right
)\mathcal{L} \left (1-\sum_{j=1}^l \Pi_j \right ),
\label{sum_project} \ee defined by the following projection
operators \be \Pi_j = \frac{W_j(k) \rangle \langle
W_j^*(k)}{\langle |W_j(k)|^2\rangle},\ \Pi_j \Pi_l =\delta_{j,l}
\Pi_j. \ee From the condition of orthogonalization of the
dynamical variables we obtain the initial values for the TCF's of
Eq. \Ref{TCF}: \be M_{jl}(k,t=0)=\left\{
\begin{array}{rcl}
0, & \textrm{if}\ j\neq l, & \textrm{cross-correlations}\\
1, & \textrm{if}\ j=l, & \textrm{autocorrelations}\\
\end{array}
\right. \ee These correlation functions $M_{jl}(k,t)$ are
symmetrical in $l$ and $j$, i.e., \be M_{jl}(k,t)=M_{lj}(k,t).
\label{sym} \ee Autocorrelation functions of Zwanzig-Mori
formalism have the following property: every autocorrelation
function of the higher order $M_j(k,t)=M_{jj}(k,t)$ is a memory
function for the previous one, i.e. $M_{j-1}(k,t)$
(autocorrelation functions will be marked by one index only in
accordance with the used variable), and they  are interrelated by
integro-differential non-Markovian equations of the form: \be
\frac{dM_{j-1}(k,t)}{dt}+ \Omega_{j}^{2}(k) \int_{0}^{t}d\tau
 M_{j}(k,\tau)M_{j-1}(k,t-\tau)=0.
\label{first_eq}
\ee
Differentiating the first equation of the chain \Ref{first_eq},
i.e. $j=1$, one obtain the generalized Langevin equation:
\begin{widetext}
\begin{equation}
\frac{d^{2}M_0(k,t)}{dt^{2}}+\Omega_{1}^{2}(k)M_0(k,t)-
\Omega_{1}^{2}(k)\Omega_{2}^{2}(k) \int_{0}^{t}d\tau
\int_{0}^{\tau}d\tau^{'}M_{2}(k,t-\tau)
M_{1}(k,t-\tau^{'})M_0(k,\tau^{'})=0\;. \label{second}
\end{equation}
\end{widetext}
One the other hand, these functions describe concrete relaxation
processes, the physical meaning of which may be established from
direct definitions of TCF's. For instance, $M_0(k,t)$ describes
the dynamics of fluctuations of density correlations in the
system, $M_1(k,t)$ is the TCF of the fluctuations of the
longitudinal component of the momentum density, $M_2(k,t)$
contains the TCF of fluctuations of energy density. So, these
quantities are associated with the TCF's of the well-known
hydrodynamic ``slow'' variables. These TCF's have characteristic
time scales, which can be found from \be \tau_{j}(k)=\textrm{Re}\;
\int_{0}^{\infty}dt \; M_{j}(k,t) = \textrm{Re}\;
\widetilde{M}_{j}(k,s=0), \label{tau} \ee where
$\widetilde{M}_{j}(k,s)$ is the Laplace transform of the
corresponding TCF, i.e.
$\widetilde{M}_{j}(k,s)=\int_{0}^{\infty}dt\; e^{-st} M_{j}(k,t)$
\cite{time,Egelstaff,Costa}.

So, the memory function approach with single initial dynamical
variable extracts the whole set, which describes the relaxation
processes of the corresponding relaxation levels. In fact, the
well-known problem of the choice of a set of variables required
for the correct description of the system dynamics here is reduced
(i) to the search of the number of variables for \textit{a priori}
known succession $\textbf{W}(k)$, that was excellently shown by
the recurrent relation approach in a works of Lee
\cite{Lee,Omega}; and/or (ii) to the finding the correct closure
of the chain \Ref{first_eq}.

The ratio between $\tau_0(k)$, $\tau_1(k)$ and $\tau_2(k)$ may be
quite arbitrary. In the hydrodynamic region ($k \to 0$, $\omega
\to 0$) they take large values due to the slow changes of the
correspondent variables: densities of mass, momentum and energy.
Further, one can suggest that the relaxation times of the
following TCF's, in comparison with the scales of these three
variables, are comparable, i.e. $\tau_3(k) \approx \tau_4(k)$. We
emphasize here that this assumption does not contradict the
viscoelastic model, which presupposes that $\tau_2(k) \gg
\tau_3(k)$. Obviously, this key condition of the viscoelastic
theory is just a special case in our approach. Simultaneously, our
approach does not deny the presence of the long-lasting time tail
of $M_2(k,t)$, which may be adequately taken into account by the
mode-coupling theory \cite{Gotze}.

Then, taking into account Eq. \Ref{tau} one can find \be
M_{4}(k,t)=M_3(k,t)+h(k,t), \label{tail} \ee where the ``tail''
function $h(k,t)$ appears. From the short-time asymptotic of the
time autocorrelation functions and the condition of the long time
attenuation of correlation Eq. \Ref{tail} yields the following
properties of $h(k,t)$: \be \lim_{t \to 0} h(k,t) = \lim_{t \to
\infty} h(k,t) =0, \ee this function must have at least one
crossing with the time axis at the intermediate region
\cite{tail}. Eq. \Ref{tail} allows us to obtain the closure of
hierarchy of equations of the form \Ref{first_eq} at the fourth
level ($j=4$) and by means of Laplace transformation to find its
exact solution for $\widetilde{M}_0(k,i\omega)$, in particular,
which is directly related to the experimentally available term,
the dynamic structure factor, $S(k,\omega)$. The expression for
the resulting $S(k,\omega)$ is given in work \cite{our_JCP} in
terms of the first four frequency parameters $\Omega_1^2(k)$,
$\Omega_2^2(k)$, $\Omega_3^2(k)$, $\Omega_4^2(k)$ and the Laplace
transform of tail function, i.e., $\widetilde{h}(k,i\omega)$. In
some cases, the regime with $h(k,t) \to 0$ may be realized. It can
be observed in some parts of time (frequency) scale. In this case
we find the following expression for the dynamic structure factor:
\begin{widetext}
\bn
\label{Basic} S(k, \omega)&=& \frac{S(k)}{2 \pi} \Omega_{1}^{2}(k)
\Omega_{2}^{2}(k) \Omega_{3}^{2}(k) [4 \Omega_{4}^{2}(k)-
\omega^{2}]^{ \frac{1}{2}} \lbrace \Omega_{1}^{4}(k)
\Omega_{3}^{4}(k)
                                                       \nonumber \\
&+&\omega^{2}[\Omega_{1}^{4}(k) \Omega_{4}^{2}(k)-2
\Omega_{1}^{2}(k) \Omega_{3}^{4}(k)- \Omega_{1}^{4}(k)
\Omega_{3}^{2}(k)\nonumber\\
& &\ \ \ \ \ +2 \Omega_{1}^{2}(k) \Omega_{2}^{2}(k)
\Omega_{4}^{2}(k)- \Omega_{1}^{2}(k) \Omega_{2}^{2}(k)
\Omega_{3}^{2}(k)+ \Omega_{2}^{4}(k) \Omega_{4}^{2}(k)]
                                                     \nonumber \\
&+&\omega^{4}[ \Omega_{3}^{4}(k)-2 \Omega_{1}^{2}(k)
\Omega_{4}^{2}(k) +2 \Omega_{1}^{2}(k) \Omega_{3}^{2}(k) -2
\Omega_{2}^{2}(k) \Omega_{4}^{2}(k)+ \Omega_{2}^{2}(k)
\Omega_{3}^{2}(k)]
\nonumber\\
&+& \omega^{6}[ \Omega_{4}^{2}(k)- \Omega_{3}^{2}(k)]
\rbrace^{-1}.
\en
\end{widetext}
This equation is also expressed through the first four frequency
parameters, which are directly related to the first five even
frequency moments of dynamics structure factor. It is necessary to
note that this expression is obtained in the way completely
different from the theory of moments \cite{Orthner}.

\subsection{\label{sec:level3} Comparison with IXS experiment and
relationship with other theoretical approaches}

In Fig. $1$ we report the dynamic structure factor $S(k,\omega)$
of liquid lithium ($T=475$K) for some wave numbers calculated from
Eq. \Ref{Basic} (solid line) and obtained from IXS experiment
(circles) \cite{Tullio1}. Being used in theoretical computations
the static structure factor $S(k)$ for both cases was taken from
Ref. \cite{Waseda}. The first frequency parameter was directly
defined from its definition $\Omega_1^2(k)=K_B T k^2/m S(k)$. The
second frequency parameter $\Omega_2^2(k)$ is related to the
fourth frequency moment. We found this parameter from the values
of the infinite frequency sound velocity $c_{\infty}(k)$
\cite{Tullio0,Tullio1} by means of relation
$c_{\infty}(k)=\sqrt{\Omega_1^2(k)+\Omega_2^2(k)}/k$. The
high-order parameters were found by comparison with the
experiment. Eventually, we have revealed that all the frequency
parameters have the similar dispersion. In particular, they have
the first principal maximum at the same wave numbers such as the
side peak of $S(k,\omega)$, i.e. at $k \sim 0.55 k_m$, and any low
order parameter is less than the high order one.

We would like to emphasize that the theoretical $S(k,\omega)$ and,
in particular, the position of the side peak, is very sensitive to
the magnitude of $\Omega_2^2(k)$. The magnitudes of
$\Omega_3^2(k)$ and $\Omega_4^2(k)$ influence the form of
$S(k,\omega)$. However, it is not so important to know these
parameters separately as their ratio, i.e.
$\Omega_4^2(k)/\Omega_3^2(k)$.

To compare the theoretical outcome with the experiment we modified
it to account for the quantum mechanical detailed balance
condition according to \be S_{q}(k,\omega) \simeq \frac{\hbar
\omega/K_B T}{1-e^{-\hbar \omega/K_B T}} S(k, \omega),
\label{dbcond.} \ee and then broadened it for the finite
experimental resolution effects $R(k,\omega)$ \cite{Tullio0}: \be
\int R(k,\omega-\omega')S_q(k,\omega') d\omega. \label{resol} \ee
From Fig. $1$ one can see that the above described theoretical
approach yields a good agreement with IXS data of both systems.

Now we can execute a more detailed study of the obtained results
and compare them with other approaches: the usual viscoelastic
model, the double-scale viscous model and the generalized mode
approach. The common feature of these theories is the use of the
time autocorrelation function $M_{jl}(k,t)$ of Eq. \Ref{TCF} at
$j=l=2$. So, the viscoelastic and the double-viscosity models are
based on approximations to this term, and $M_2(k,t)$ plays a key
role in these theories. As for our approach, it gives the
following form for Laplace transform of $M_2(k,t)$
\begin{subequations}
\label{eq:whole} \bn \widetilde{M}_{2}(k,s)&=&[s+
\Omega_{3}^{2}(k) \widetilde{M}_{3}(k,s)]^{-1}\nonumber\\ &=&
\displaystyle{\frac{s+\Omega_{4}^{2}(k) \widetilde{M}_{3}(k,s)}
{s^{2}+ \Omega_{4}^{2}(k)\widetilde{M}_{3}(k,s)s+
\Omega_{3}^{2}(k)}},\label{M2:1} \en \be
\widetilde{M}_{3}(k,s)=\displaystyle{\frac{-s +\sqrt{s^{2}
+4\Omega_{4}^{2}(k)}}{2\Omega_{4}^{2}(k)}}, \label{M2:2} \ee
\end{subequations}
which are obtained by Laplace transform of the third and fourth
($j=3,4$) equations of the chain \Ref{first_eq}.

To pass from the frequency dependence of
$\widetilde{M}_2(k,i\omega)$ to the time one, let us consider the
low frequency region restricted by the value $2\Omega_{4}(k)$. For
convenience we introduce here a small parameter (at the fixed wave
number $k$): \be \xi=\frac{s^{2}}{4\Omega_{4}^{2}},\ |\xi| \ll 1.
\ee Taking into account the fact that the found values of
$\Omega_{4}^{2}(k)$ for liquid sodium and lithium achieve
$10^{29}-10^{30} s^{-2}$ for the low-$k$ region, we span by
introducing parameter $\xi$ the frequency (time) range $\omega <
10^{15}s^{-1}$ ($t>10^{-15} s$), which is important for us and is
available experimentally.

Expanding the radicand in Eq. \Ref{M2:2} as a series in the
parameter $\xi$ \be
\sqrt{1+\xi}=1+\frac{\xi}{2}-\frac{\xi^{2}}{8}+...\ ,
\label{Series} \ee we can rewrite it in the following way \be
\widetilde{M}_{3}(s)=-\frac{s}{2\Omega_{4}^{2}}+\frac{1}{\Omega_{4}}
+\frac{s^2}{8\Omega_{4}^3}-\frac{s^{4}}{32\Omega_{4}^5}+...\ .
\label{second_ser} \ee By restricting the number of terms in the
series \Ref{Series} [and, accordingly, in Eq. \Ref{second_ser}] we
receive from Eq. \Ref{M2:1} the linear combination of the Lorentz
functions \be \widetilde{M}_2(k,s)=\sum_{j}
\frac{A_{j}(k)}{s+\tau_{j}^{-1}(k)}, \ j=1,2,3,5,\ldots,
\label{Lor} \ee the number of which will be increased at the
increase of the number of terms in the series \Ref{Series}. The
quantities $A_j(k)$ and $\tau_j(k)$ are expressed by the
relaxation frequencies $\Omega_3^2(k)$ and $\Omega_4^2(k)$. Going
over to the time scale by the inverse Laplace transform
\cite{Abramov} we obtain \be M_{2}(k,t)=\sum_{j}
A_{j}(k)e^{-t/\tau_{j}(k)}. \label{rjad} \ee By restricting the
first term of the series \Ref{Series} only we receive the simplest
model from the first equality of Eq. \Ref{M2:1} with Eq.
\Ref{second_ser} \be M_2(k,t)=e^{-t/\tau(k)}, \label{viscoel} \ee
which corresponds to the viscoelastic model with the relaxation
time $\tau(k)=\Omega_{4}(k)/\Omega_{3}^{2}(k)$, and from the
second equality of Eq. \Ref{M2:1} the double exponential model,
i.e Eq. \Ref{rjad} at $j=2$, with the following time relaxation
parameters \bn \tau_{1,2}(k)= \left [ \Omega_{4}(k) \pm
\sqrt{\Omega_{4}^{2}(k)- \Omega_{3}^{2}(k)} \right ]^{-1} \en and
the weight factor \be
A(k)=\frac{\Omega_{4}(k)+\sqrt{\Omega_{4}^{2}(k)-
\Omega_{3}^{2}(k)}}{2\sqrt{\Omega_{4}^{2}(k) -\Omega_{3}^{2}(k)}}.
\ee This case may be related to the double-time viscous model
\cite{Tullio1,Tullio0}, two-time exponential \textit{ansatz}
\cite{Levesque,Egelstaff}. In the general form Eq. \Ref{rjad}
corresponds to the framework of generalized collective mode
approach \cite{Schepper} with the sum of the weighed exponents for
the TCF $M_2(k,t)$, where $\tau_j^{-1}(k)$ denote eigenvalues of a
generalized dynamic matrix with the elements consisting of static
correlation functions, and the weight factors $A_j(k)$ are the
amplitudes describing the contribution of the corresponding modes.

So, it is obvious that the theory underlying Eq. \Ref{Basic}
prescribes such behavior of the second order memory function
$M_2(k,t)$, which may be represented in the form of Eq. \Ref{rjad}
and can be reduced to the above-mentioned models.  Eq. \Ref{rjad}
is in fact an expansion of $M_{2}(k,t)$ into decay channels
embedded in this function.

\section{Scale uniformity of dynamics processes in liquid alkali metals}
\label{scale}

The determination of the scale uniformity of structural and
dynamical features for different groups of liquids is very
important for the physics of liquid state. On the one hand, it
allows one to apply the unified theoretical description to the
whole group. On the other hand, it allows one to remove the
difficulties related to obtaining the experimental data. The fact
is that until recently the microscopic dynamics of liquids could
be experimentally probed by INS only. However, there were often
different problems related to, firstly, separation of collective
and one-particle contributions, and, secondly, gross experimental
errors (and even with impossibility to obtain data) for different
($k,\omega)$-regions. Recent progress in the technique of IXS has
allowed one to clear some of the obstacles \cite{Burkel}. Ten
years ago the possibility of the unified description of the
structural and dynamical properties of different liquid alkali
metals near the melting point was found by the comprehensive
molecular dynamics simulation study \cite{Balucani_PRB}, where the
adopted potential model of Price, Singwi and Tosi was used, and
the scale passage was executed on the basis of the potential
parameters. The recent sketchy attempt of testing this outcome
experimentally has shown its inconsistency \cite{Tullio1}.

In present work we also execute the comparison of the dynamic
structure factor spectra of liquid lithium and sodium. As known
from the experimental results, the dynamic structure factor
$S(k,\omega)$ depends strongly on the temperature $T$ and the wave
number $k$. So, one can define the reduced forms of these terms as
$T/T_m$ and $k/k_m$, where $T_m$ is the melting temperature and
$k_m$ is the main peak position in the static structure factor
$S(k)$ for the corresponding system. The scale time interval $t^*$
can be expressed as $t^*=k^{-1}\sqrt{m/K_B T}$. Thus defined time
unit $t^*$ is different from the one introduced in Ref.
\cite{Tullio1}, because the present term varies with the change of
space and temperature characteristics. Although we do not exclude
the possibility, that this scale unit may be independent of the
temperature and the wave number for other systems (for instance,
semi-conductors, or H-bonded liquids).

In Fig. $2$ we report the comparison of $S(k,\omega)$ spectra for
liquid lithium and sodium \cite{Tullio0,Tullio1} at approximately
the same reduced temperatures $T/T_m$ and wave numbers $k/k_m$.
Namely, $T/T_m=1.049$ for liquid lithium and $1.051$ in case of
sodium.
From this figure one can see that dynamic structure factor
practically coincides in the first two higher cases. From the
lower plot of Fig. $2$ one can see, that the position of inelastic
and central peaks for both systems is the same. However, though
the overall coincidence of spectra is observed at intermediate
frequencies only, the peak altitudes are a little different. Such
deviation can easily be explained by the fact that the plot for
liquid lithium is presented for a higher value of the reduced wave
number, $0.75$, whereas in case of sodium $k/k_m=0.73$. As known,
these wave numbers correspond to the so-called the de Gennes
narrowing region characterized by a strong $k-$dependence. In
other words, a higher section at $k$ of flat $S(k,\omega)/t^*$ is
presented for lithium than for sodium. It is necessary to take
into account, that the values of the reduced temperatures for both
systems are also slightly different.

Notice that the time unit $t^*$ depends on the system features
($m$), on the probed spatial region ($k$) and the temperature
regime ($T$) in contrast to scale units $k_m$ and $T_m$, which
remain unchanged the spatial region and the temperature of the
system is revised.

At result, the experimental or theoretical $S(k,\omega)$ for any
single metal allows one easily to restore this term for the whole
group of alkali metals at same reduced conditions, $k/k_m$ and
$T/T_m$. Moreover, the theory developed for the concrete separate
alkali metal may be simply extended to the whole group.

As an example, in Fig. $3$ we report the dynamic structure factor
of liquid potassium $S^K(k,\omega)$ obtained on the basis IXS data
for liquid sodium $S^{Na}(k,\omega)$. The transition
$S^{Na}(k,\omega) \rightarrow S^K(k,\omega)$ has been executed by
means of the following scale reductions:
\begin{subequations}
\label{eq:whole}
\begin{equation}
S^K(k,\omega)= S^{Na}(k,\omega)\frac{k^{Na}}{k^{K}}\sqrt{\frac{m^K
T^{Na}}{m^{Na} T^K}}, \label{scale:1}
\end{equation}
\begin{equation}
\omega^{K}=\omega^{Na} \frac{k^{K}}{k^{Na}} \sqrt{\frac{m^{Na}
T^K}{m^K T^{Na}}}, \label{scale:2}
\end{equation}
\begin{equation}
T^K=\frac{T_m^K T^{Na}}{T_m^{Na}},\; k^K=\frac{k_m^K
k^{Na}}{k_m^{Na}}. \label{scale:3}
\end{equation}
\end{subequations}
By the top subscript we note the corresponding system ($K$ or
$Na$).

\section{Concluding remarks} \label{remarks}

The following results are presented in this work.

(i) The theory, developed on the basis of Bogoliubov ideas about
the hierarchy of relaxation times, allows one to obtain dynamic
structure factor, reproducing adequately experimental IXS spectra
for liquid alkali metals (in particular, for liquid lithium and
sodium) in the region of low values of wave number.

(ii) The expansion of the second order memory function into
exponential decay channels, used (sometimes intuitively) in others
theories, may be easily obtained within the framework of the
presented approach. This is the evidence of the multi-mode
character of decay of the observed relaxation process.

(iii) An important result of this work is the confirmation of the
proposition about the unitary description of the dynamical
features of liquid alkali metals, and finding of corresponding
scale transition relations.

\section{Acknowledgments} The authors are
grateful to T. Scopigno for providing IXS data, and to M. H. Lee
for fruitful discussions. A. V. M. acknowledges G. Garberoglio and
F. A. Oliveira for the useful correspondence. This work was
supported by the Russian Ministry of Education and Science (Grants
No. 03-06-00218a, A03-2.9-336) and RFBR (No. 02-02-16146).

\newpage

\vspace{0.5cm}

\section{Figure captions}

Fig. 1. Dynamic structure factor of liquid lithium at the
temperature $T=475$K. The solid lines are the results of the
theoretic model \Ref{Basic}, whereas the open circles are the IXS
data \cite{Tullio0}. The theoretical lineshapes have been modified
to account for the quantum mechanical detailed balance condition
and broadened for the finite experimental resolution effects as
described in the text. The wave numbers $k$ are given in a reduced
form, where $k_m$ is the main peak position in the static
structure factor $S(k)$.

Fig. 2. IXS spectra of liquid lithium at $T=475$K ($\Box \Box
\Box$) and liquid sodium at $T=390$K ($\circ \circ \circ$)
\cite{Tullio0,Tullio1} in the reduced units. The scale frequency
$\omega^*$ is chosen as the term inverse proportional to $t^*$.

Fig. 3. The dynamic structure factor $S(k,\omega)$ of liquid
potassium at $T=354.1$K calculated from IXS data of liquid sodium
at $T=390$K by the scale reduction described in the text.

\newpage

\begin{figure}
\centering
\includegraphics[width=.48\textwidth]{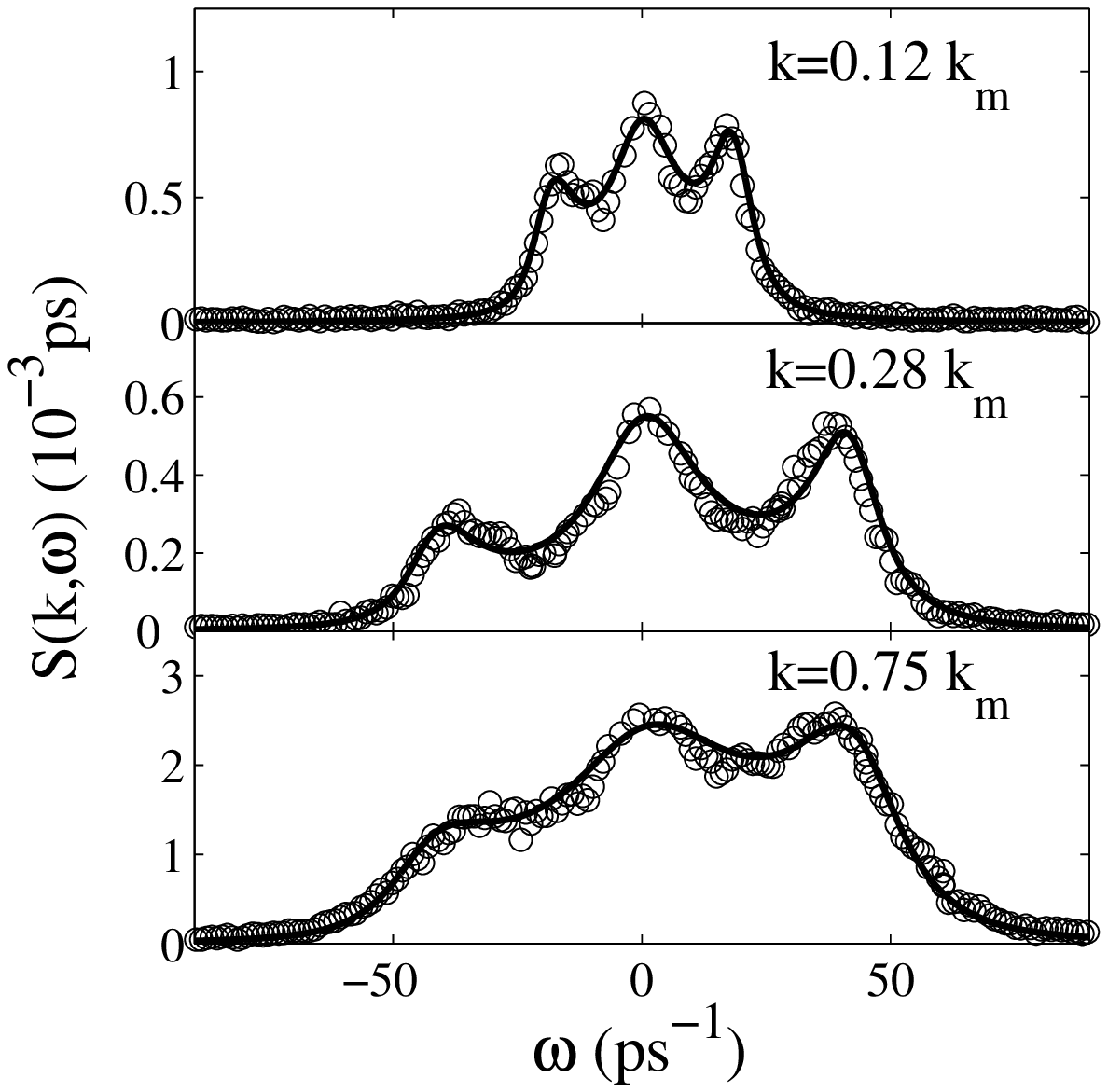}
\label{Skw_Li_Na}
\end{figure}

\vspace{7.cm} Fig. 1
\newpage

\begin{figure}
\centering
\includegraphics[width=.48\textwidth]{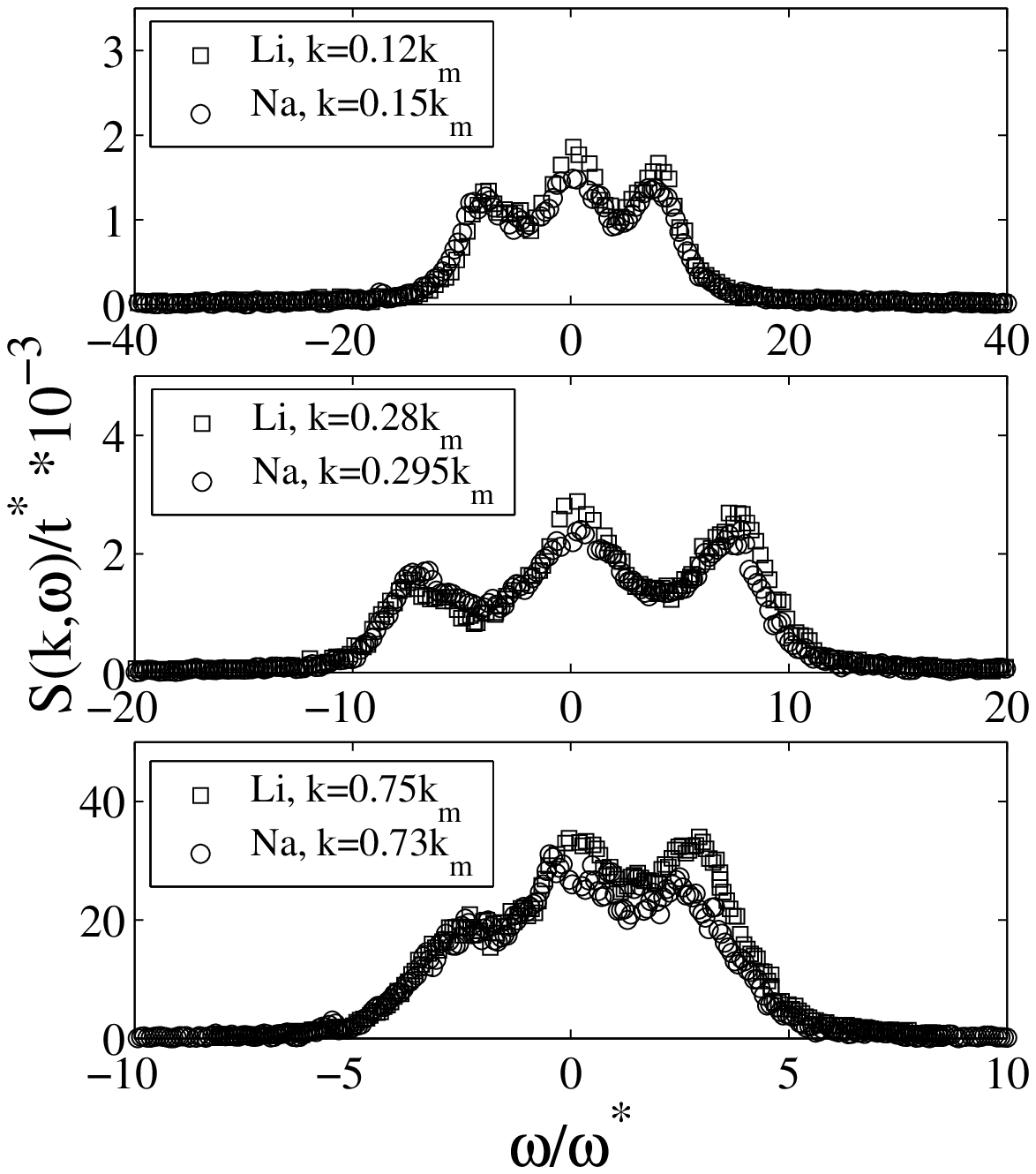}
\label{univ}
\end{figure}

\vspace{7.cm} Fig. 2 
\newpage

\begin{figure}
\centering
\includegraphics[width=.48\textwidth]{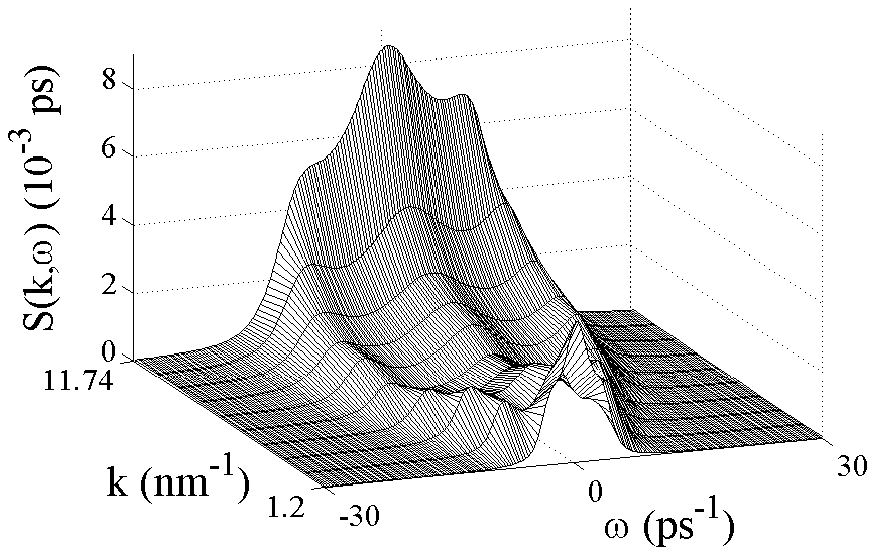}
\label{potassium}
\end{figure}

\vspace{7.cm} Fig. 3 


\begin{thebibliography}{99}
\bibitem{time} Eq. \Ref{tau} defines square under the function $M_j(k,t)$. It
takes large values (is turned to infinity) for slow damping
(undamping) processes, and it is finite for ordinary ergodic
cases. So, it can be chosen as an estimation measure of the
relaxation scale. The main advantage of such definition consists
in direct establishing relation to the experiment. For example,
$\tau_0(k)=\pi S(k,\omega)/S(k)$. There are other definitions of
relaxation time. For instance, P. Egelstaff and W. Gl\"aser have
shown in Ref. \cite{Egelstaff} that ``time of TCF'' may be found
from $\tau_j(k) = \int_{0}^{\infty} t \; M_j(k,t)\;
dt/\int_{0}^{\infty} M_j(k,t)\; dt$. Some ideas about relaxation
time are also given in Ref. \cite{Costa}.
\bibitem{Omega} Via recurrent relations approach was shown that
finding, for example, of $M_0(k,t)$ can be reduced to the
definition of the dimension of a space formed by the set of basic
``vectors'' $W_0$, $W_1$, $W_2$, $\ldots$, $W_j$. For infinite
spaces at $j \to \infty$ given TCF may be also found at the known
parameters $\Omega_1^2(k)$, $\Omega_2^2(k)$, $\ldots$ or at the
known rule of their intercoupling. Taking into account the fact
that $\Omega_j^2(k)$ contains $j$th-particle distribution
functions, the quest for their intercoupling may be reduced to the
quest for the way to uncouple many-particle distribution
functions. However, even on solving this problem the quest for the
final result is complicated by multiplex reckonings.
\bibitem{tail} It should not be confused with the tail function of MCT, where
it defines the long-time behavior of relaxation, whereas the
function $h(k,t)$ is the difference of $M_4(k,t)$ and $M_3(k,t)$.
\end{thebibliography}
\end{document}